\documentstyle[twocolumn,aps,prl]{revtex}

\input epsf

\begin{document}

\title{Black Hole Criticality in the Brans-Dicke Model}
\author{Steven L. Liebling and Matthew W. Choptuik}
\address{Center for Relativity, The University of Texas, 
         Austin, TX 78712-1081}

\date{\today}

\maketitle

\begin{abstract}
We study the collapse of a free scalar field in the
Brans-Dicke model of gravity. At the critical point
of black hole formation, the model admits two distinctive
solutions dependent on the value of the coupling
parameter.  We find one solution to be discretely
self-similar and the other to exhibit
continuous self-similarity. 
\end{abstract}

\pacs{04.25.Dm, 04.40.-b, 04.70.Bw}
   
Studies of black hole formation from the gravitational collapse of 
a massless scalar field have revealed interesting nonlinear phenomena
at the threshold of black hole formation \cite{matt,EHH}. These
studies have shown that Einstein's field equations possess
solutions which occur precisely at the black hole threshold and
which are universal with respect to the initial conditions of
the evolution.
More specifically, for any type of initial field configuration 
whose energy is parameterized 
by some  parameter, $p$, the critical solution occurs at
a value of $p=p^*$ such that for all $p<p^*$ no black hole is
formed, and for all $p>p^*$ a black hole is necessarily formed. This
critical solution, whether obtained with an initial pulse shape such as
$\tanh$ or a Gaussian pulse, is identical, erasing all detail
of the initial field configuration.

Though universal with respect to initial conditions, the
critical solution is dependent on the specific matter model
involved. In the case of a real scalar field \cite{matt}, 
a discretely self-similar solution~(DSS)
is found, characterized by an echoing exponent $\Delta$.
In other words, were an observer to take a snap-shot of 
the solution at some time $t$, he would find the same picture 
as when he zoomed in to a spatial scale $\exp(\Delta)$ smaller 
than the original at a time $t+\exp(\Delta)$ later. 

In contrast to this DSS solution, other researchers,
working in an axion/dilaton model, have found that the
equations possess a continuously self-similar (CSS) solution \cite{EHH}. 
Because they found this solution by assuming continuous self-similarity
and solving the appropriate ordinary differential equations, they could
not show whether this CSS solution is indeed a critical solution. 

We find that a free real scalar field coupled to Brans-Dicke
gravity contains two distinct dynamic critical solutions.
As a special case, the model includes the real scalar field
in general relativity and recovers the DSS solution as in \cite{matt}.
Further, this model is sufficiently general that it contains the model
studied in \cite{EHH} as another special case. For this case, we
find that the CSS solution is an attracting critical solution.
Hence we present the novel result that for a single matter model,
adjustment of a coupling parameter transitions between two unique, dynamic,
attracting critical solutions. Because these two solutions are
both dynamic, the model is quite different from  the Yang Mills
model studied in \cite{matt2}.

Subsequent to our study, Hirschmann and Eardley, working in
an even more general model, the non-linear sigma model,
 which includes ours, carry-out a
perturbation analysis and confirm a change
in stability near the value we find for the transition
coupling parameter \cite{he}.  Further, from the eigenvalues
of the unstable modes, they have  been able to compute mass-scaling
exponents. Their results concur with those we find from our numerical
evolutions.

We work in spherical symmetry with the metric
\begin{equation}
ds^2  = -\alpha(r,t)^2 dt^2 + a(r,t)^2 dr^2 + r^2 d\Omega^2,
\end{equation}
where $\alpha(r,t)$ represents the lapse function in the 3+1
formalism and $r$ measures proper surface area.

The Brans-Dicke model is described by the field equations
\begin{equation}
G_{\mu \nu} = \frac{8 \pi}{\phi(r,t)} T^{\rm total}_{\mu \nu}
\end{equation}
where $1/\phi(r,t)$ represents the freedom of the conventional 
gravitational constant to vary \cite{bd}. The total stress-energy tensor
consists of two terms
\begin{equation}T^{\rm total}= T^{\rm matter} + T^{\rm BD},
\end{equation}
where $T^{\rm BD}$ represents the energy associated with the Brans-Dicke
field $\phi$ and $T^{\rm matter}$ is the conventional tensor
associated with matter sources \cite{bd}. For this study our sole 
matter source is
a free massless minimally coupled scalar field $\psi(r,t)$ governed by
\begin{equation}
\Box \psi = 0
\end{equation}
and whose associated stress-energy is 
\begin{equation}
T^{\rm matter}_{\mu\nu} =  \psi_{,\mu} \psi_{,\nu} -
        \frac{1}{2} g_{\mu\nu}  \psi^{,\rho} \psi_{,\rho}.
\end{equation}
The Brans-Dicke field satisfies the generally covariant wave equation
\begin{equation}
\Box \phi = 4\pi \lambda T^{\rm matter}
\end{equation}
where $\lambda$, a constant, represents the strength of the coupling
between the Brans-Dicke field and matter \cite{weinberg}. 
Its associated stress-energy
tensor is
\begin{equation}
T^{\phi}_{\mu \nu} =\frac{\omega}{8 \pi \phi} \left(
        \phi_{,\mu} \phi_{,\nu} -\frac{1}{2} g_{\mu \nu}
        \phi_{,\rho} \phi^{,\rho} \right) +
        \frac{1}{8\pi} \left(\phi_{,\mu \nu} -
        g_{\mu \nu} \Box \phi \right)
\end{equation}
where
\begin{equation}
\lambda \equiv \frac{2}{2 \omega + 3}.
\end{equation}
The equations described above are said to be in the {\it
Brans-Dicke frame} where masses are constant but inertial
forces depend on the distribution of mass in the universe.
However, it is possible to transform to a conformal frame in which
the geometry is described by Einstein's field equations with
vanishing second derivatives of $\phi$.
In this frame, the {\it Einstein frame}, masses vary with time, but
the gravitational constant is indeed constant.

We achieve this conformal transformation via
\begin{eqnarray}
        e^{\xi} & \equiv & \phi \nonumber \\ 
	g_{\mu \nu} & \rightarrow  & e^{\xi} g_{\mu \nu}  \\
	g^{\mu \nu} & \rightarrow & e^{-\xi}g^{\mu \nu}  \nonumber 
\end{eqnarray}
after which we have the equations (now expressed in the Einstein
frame)
\begin{eqnarray}
T^{\rm BD} & = & \left( \frac{3+2\omega}{16 \pi} \right) \left(\xi_{,\mu}
	\xi_{,\nu}- \frac{1}{2}  g_{\mu \nu}  \xi^{,\rho}
	\xi_{,\rho} \right)    \\
T^{\rm matter}_{\mu \nu} & \rightarrow & \frac{1}{\phi} T^{matter}_{\mu \nu}
                \\
\Box \xi  & = & -4\pi \lambda e^{-\xi} \psi^{,\mu} \psi_{,\mu}
                 \\
\Box \psi & = & \psi^{,\mu} \xi_{,\mu}.
\end{eqnarray}

We define auxiliary variables in terms of the derivatives of the
scalar fields
\begin{eqnarray}
\Phi_\xi   \equiv  \frac{\partial}{\partial r} \xi   & {\rm ~~~and~~~} &
\Pi_\xi    \equiv  \frac{a}{\alpha} \frac{\partial}{\partial t}  \xi   \\
\Phi_\psi  \equiv  \frac{\partial}{\partial r} \psi & {\rm ~~~and~~~} &
\Pi_\psi   \equiv  \frac{a}{\alpha} \frac{\partial}{\partial t}  \psi  
\end{eqnarray}
so that the wave equations result in 
\begin{eqnarray}
\dot \Phi_\xi & = & \left( \frac{\alpha}{a} \Pi_\xi \right)'  \nonumber \\
\dot \Pi_\xi & = & \frac{1}{r^2} \left( \frac{r^2 \alpha}{a} \Phi_\xi \right) ' +
        4\pi \lambda e^{-\xi}\frac{\alpha}{a} \left( \Phi_\psi^2 -
        \Pi_\psi^2 \right)  \label{eq:eom}\\
\dot \Phi_\psi& = & \left( \frac{\alpha}{a} \Pi_\psi\right)'  \nonumber \\
\dot \Pi_\psi & = & \frac{1}{r^2} \left(\frac{r^2 \alpha}{a} \Phi_\psi\right) ' +
        \frac{\alpha}{a} \left( \Pi_\psi\Pi_\xi -\Phi_\psi\Phi_\xi \right). 
	\nonumber 
\end{eqnarray}
The only other necessary conditions come from the field equations, which,
in the Einstein frame, are simply Einstein's field equations, 
$G_{\mu \nu} = 8 \pi T_{\mu \nu}$.  In accordance
with the 3+1 formalism, we have the Hamiltonian constraint 
\begin{equation}
a'        =   - \frac{a^3-a}{2r} 
                + 2 \pi a r \left( e^{-\xi} \left( \Phi_\psi^2 + 
                \Pi_\psi^2 \right) + \frac{1}{8\pi \lambda} \left( \Phi_\xi^2 + 
                \Pi_\xi^2  \right) \right) 
\label{eq:hamiltonian} 
\end{equation}
and the polar slicing condition
\begin{equation}
\alpha'   =  -\left( \frac{1-a^2}{r} 
                - \frac{a'}{a} \right) \alpha,
\label{eq:slicing}
\end{equation}
which enable us to solve for the geometry in terms of the two 
sources, $\psi$ and $\xi$.
These equations suffice to evolve both the fields $\psi(r,t)$ and $\xi(r,t)$,
and the geometric variables $\alpha(r,t)$ and $a(r,t)$ \cite{steve}.

To show that the model found in \cite{EHH} is a special case of our
model, we compare our Lagrangian
\begin{equation}
L^{\rm BD} = -\frac{1}{2} e^{-\xi} \psi^{,\rho} \psi_{,\rho}
                - \frac{1}{16\pi \lambda} \xi^{,\rho} \xi_{,\rho}
\label{eq:lbd}
\end{equation}
with that of \cite{EHH}
\begin{equation}
L^\tau = -\frac{1}{32\pi} \left(
        e^{4\phi} a_{,\mu} a^{,\mu} + 4 \phi_{,\mu} \phi^{,\mu}
                        \right),
\label{eq:ltau}
\end{equation}
defined in terms of the axion, $a$, and the dilaton, $\phi$.
Comparing Eqs.~(\ref{eq:lbd}) and (\ref{eq:ltau}), we see
a correspondence between the two models with a trivial
rescaling of the fields
\begin{equation}
\xi     =  -4 \phi   {\rm ~~~~~} \psi     =  \frac{1}{\sqrt{16\pi}} a  
{\rm ~~~~~} \lambda  =  8.    \label{eq:correspond}
\end{equation}

\begin{figure}
\epsfxsize=7cm
\centerline{\epsffile{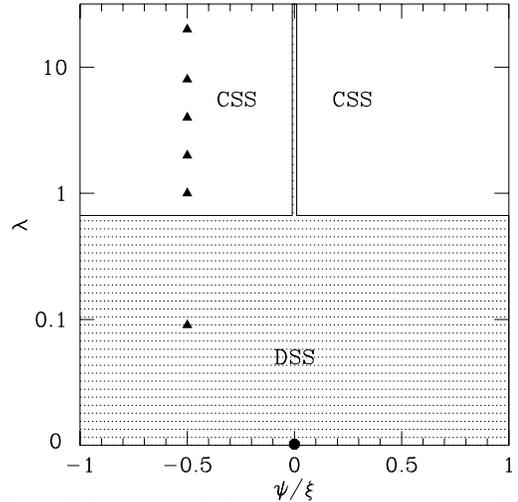}}
\caption{Schematic of the configuration space. The horizontal
axis represents the mixing ratio between the free scalar field
and the Brans-Dicke field.  The darkened triangles represent
the locations of the profiles displayed in Fig. 2. The darkened
circle represents the location of the real scalar field in
general relativity studied in [1].
For $\lambda < 2/3$, only
the DSS solution is the attractor. Above $\lambda \approx 2/3$ the
CSS solution attracts whenever  both fields are initially present.}
\label{FIG1}
\end{figure}

We have found the critical solutions for a variety of initial
data. Specifically, we input the initial configuration of the
two fields, and specify the value of $\lambda$. The space of
initial configurations is schematically represented 
in Fig.~\ref{FIG1}.

%%%%%%%%%%%%%%%%%%%%%%%%%%%%%%%%%%%%%%%%%%%%%%%%%%%%%%%%%%%%%%%%%%%%%%
%%%%%%%%%%%%%%%%%%%%%%%%%%%%%%%%%%%%%%%%%%%%%%%%%%%%%%%%%%%%%%%%%%%%%%

We observe for an initially vanishing scalar field that 
Eqs.~(\ref{eq:eom}, \ref{eq:hamiltonian}, \ref{eq:slicing})
describe the real scalar field case studied in \cite{matt}.
Consistent with this observation, our results recover the same
DSS solution found for the real scalar field case.
The equivalence between this model with $\psi(r,t)=0$ and that
of the real scalar field is shown in Fig.~\ref{FIG1}
as the vertical line extending through the middle of the graph.

When $\lambda \rightarrow 0+$, Weinberg shows that the 
Brans-Dicke model goes over to general relativity. Hence,
for the general situation in which both fields are present
($\psi/\xi \neq 0$), we expect to recover the results from
general relativity. We do recover the general relativity result,
that being the DSS solution. As shown in Fig.~\ref{FIG1},
the critical solution is discrete for generic initial data as
$\lambda$ is increased up to $\lambda \approx 2/3$.

\begin{figure}
\epsfxsize=7cm
\centerline{\epsffile{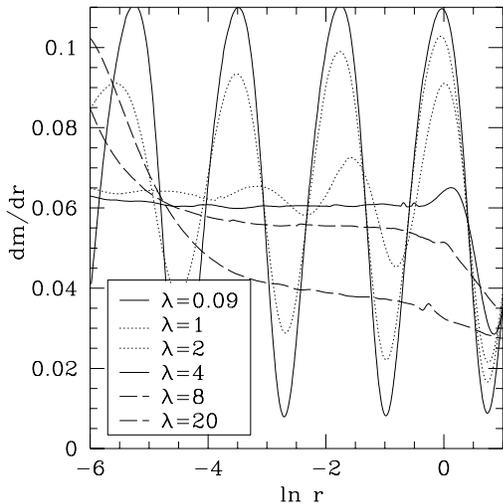}}
\caption{Demonstration of regime in which the solution
transitions from the DSS to the CSS.  We show here
the mid-section of the seven
solutions for various $\lambda$. For $\lambda=0.09$ the solution
is clearly the DSS, however the next solution demonstrates
that the echoes get damped as one moves towards the origin. Eventually
the solutions become the CSS.}
%this graph made with supermongo in :
% /usr2/people/steve/papers/eda/self_sim_families
\label{FIG2}
\end{figure}

Shown in Fig.~\ref{FIG2} for $\lambda = 0.09$, we have verified that
this is the same DSS solution obtained for the real scalar field
in general relativity \cite{matt}.  In Table \ref{tab:deltas},
we show the computed values of the echoing exponent $\Delta$. These
values correspond to that found in \cite{matt}.

%%%%%%%%%%%%%%%%%%%%%%%%%%%%%%%%%%%%%%%%%%%%%%%%%%%%%%%%%%%%%%%%%%%%%%
%%%%%%%%%%%%%%%%%%%%%%%%%%%%%%%%%%%%%%%%%%%%%%%%%%%%%%%%%%%%%%%%%%%%%%

Around $\lambda \approx 2/3$, a remarkable transition occurs in
the critical solution.  As one increases $\lambda$ in this region,
the echos displayed by the critical solution are damped by a decreasing
envelope as shown in Fig.~\ref{FIG2}.  

\begin{figure}
\epsfxsize=7cm
\centerline{\epsffile{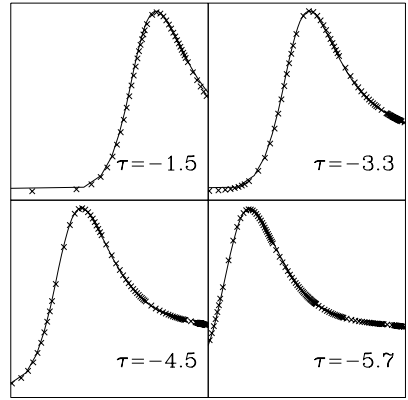}}
\caption{Demonstration of equivalence between the
CSS solution obtained from [2]
and our critical solution obtained with $\lambda = 8$.
The solid line shows the 
metric function $a(r,t)$ versus $\ln r$ provided by Eardley. The
crosses denote data points from our solution.  Four time profiles
are shown with 
$\tau=\ln(T-T^*)$, where $T$ is the central proper time of the slice
and $T^*$ is the critical time of collapse.  The
Eardley and Hirschmann solution is scaled to match our profile at $\tau=-3.3$.
The congruence at other times displays the equivalence of the two
solutions.}
%Subsequent scalings were performed by $r \rightarrow r*(T-T^*)/\exp(-3.3)$.}
%this graph produced by supermongo in directory:
%    /usr2/people/steve/papers/eda/sm/show_ehh
\label{FIG3}
\end{figure}

At $\lambda = 8$,
we recover, as expected from Eq.~(\ref{eq:correspond}),
the CSS solution found in \cite{EHH}. In Fig.~\ref{FIG3}
we demonstrate that
the solution found by \cite{EHH} by demanding continuous
self-similarity is indeed the attracting critical solution. Here
we show that by a trivial rescaling of the fields at one time slice,
our solution is identical to theirs.

%%%%%%%%%%%%%%%%%%%%%%%%%%%%%%%%%%%%%%%%%%%%%%%%%%%%%%%%%%%%%%%%%%%%%%
%%%%%%%%%%%%%%%%%%%%%%%%%%%%%%%%%%%%%%%%%%%%%%%%%%%%%%%%%%%%%%%%%%%%%%

%%%%%%%%%%%%%%%%%%%%%%%%%%%%%%%%%%%%%%%%%%%%%%%%%%%%%%%%%%%%%%%%%%%%%%
\begin{figure}
\epsfxsize=7cm
\centerline{\epsffile{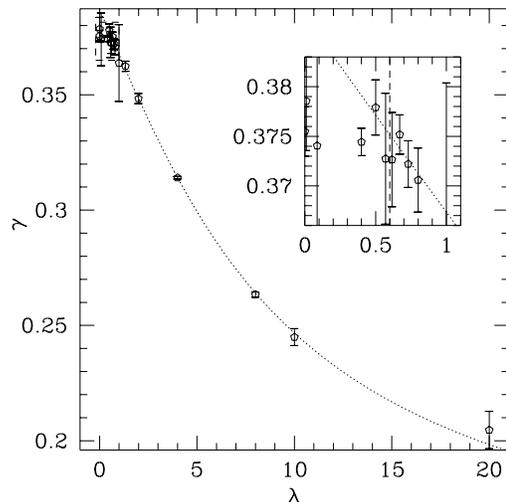}}
\caption{Black hole mass scaling exponents for various $\lambda$.
The dotted line displays the values obtained for the non-linear sigma
model in [4]. The open pentagons represent the scaling exponent
obtained by least-squares fits using our numerical results. The
errorbars represent a range of three standard deviations.}
\label{FIG4}
\end{figure}
%%%%%%%%%%%%%%%%%%%%%%%%%%%%%%%%%%%%%%%%%%%%%%%%%%%%%%%%%%%%%%%%%%%%%%

%%%%%%%%%%%%%%%%%%%%%%%%%%%%%%%%%%%%%%%%%%%%%%%%%%%%%%%%%%%%%%%%%%%%%%
\begin{figure}
\epsfxsize=7cm
\centerline{\epsffile{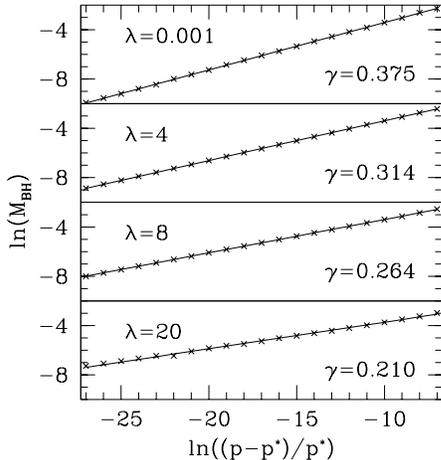}}
\caption{Illustration of the power-law mass scaling relation. The
markers display the mass obtained for the normalized distance from
criticality.  The lines designate the least-squares fit line with
slope $\gamma$.}
\label{FIG5}
\end{figure}
%%%%%%%%%%%%%%%%%%%%%%%%%%%%%%%%%%%%%%%%%%%%%%%%%%%%%%%%%%%%%%%%%%%%%%

Both the critical solutions exhibit mass-scaling in the
supercritical regime.
Specifically, for some family of solutions where
$p^*$ represents the critical value of a parameter, the masses
of the black holes formed in the regime where $p>p^*$ follow
\begin{equation}
M_{\rm BH} = c(p-p^*)^\gamma
\end{equation}
where $\gamma$ depends on $\lambda$ 
and $c$ is a family dependent constant. 
Fig.~\ref{FIG5} shows  four power law fits and 
the associated $\gamma$'s.

In keeping with the correspondence between this model
for very small $\lambda$ and that of general relativity,
we find $\gamma = 0.37$, matching that found in \cite{matt}.
Likewise, we find agreement between our values of $\gamma$
and those found by perturbation analysis in \cite{he}.
We display both these sets of values in Fig.~\ref{FIG4}.

The appearance of these two disparate solutions leads one 
to examine the transition in $\lambda$-space from the DSS 
to the CSS.  Bracketing solutions have shown that around 
$\lambda=2/3$ the transition occurs (see Fig.~\ref{FIG2}).  
As $\lambda$ is 
increased around this transition value, an envelope dampens 
the discrete echos into the smoothly continuous 
self-similar solution.  Perturbation results in the non-linear
sigma model confirm a change in stability of the CSS solution
near $\lambda=2/3$ \cite{he}.

Further parameter surveys are needed to further specify the transition
point between the two self-similar solutions. 
We also anticipate interesting {\it negative mass}
solutions for $\lambda < 0$. However, our studies have clearly shown the
richness of the solution space for even a simple, two-scalar field,
one-dimensional problem such as this one.

We are grateful to Douglas Eardley and Eric Hirschmann for 
discussions and providing us their data. 
NSF Grants PHY9310083 and PHY9318152 helped support 
this research, along with a Cray Research Grant.
Computations were performed on the facilities at the Center
for High Performance Computing at the University of Texas System.

\begin{table}
\caption{Mass scaling exponents $\gamma$ and the 
         spatial scaling exponents $\Delta$ for the various
         discretely self-similar solutions found.}
\label{tab:deltas}
\begin{tabular}{dddd}
$\lambda$ & $\omega$ & $\gamma$ & $\Delta$ \\
\tableline
0.001  &  1000    &  0.375  & \\
0.01   &  100     &  0.379  & \\
0.087  &  10      &  0.374  & \\
0.4    &  1.0     &  0.374  & \\
0.5    &  0.5     &  0.378  & \\
0.57   &  0.25    &  0.373  & 3.447\\
0.615  &  0.125   &  0.373  & \\
0.67   &  0.0     &  0.375  & 3.447 \\
0.73   &  -0.125  &  0.372  & \\
0.8    &  -0.25   &  0.371  & \\
1.0    &  -0.50   &  0.364  & \\
1.33   &  -0.75   &  0.362  & \\
2.0    &  -1.00   &  0.348  & \\
4.0    &  -1.25   &  0.314  & \\
8.0    &  -1.375  &  0.263  & \\
10     &  -1.40   &  0.245  & \\
20     &  -1.45   &  0.205  &
\end{tabular}
\end{table}


\begin{references}
\bibitem{matt} 
        M.~W.~Choptuik, 
        Phys. Rev. Lett, {\bf 70}, 9-12 (1993).
\bibitem{EHH}  
        D.~Eardley, E.~Hirschmann and J.~Horne,
        Phys. Rev.  {\bf D52}, 5397-5401 (1995).
\bibitem{matt2} M.~W.~Choptuik, T.~Chmaj, and P.~Bizo\'n. LANL
        preprint gr-qc/9603051 (1996).
\bibitem{he} E.~W.~Hirschmann and D.~M.~Eardley,
        LANL preprint gr-qc/9511052 (1995).
\bibitem{bd} C.~Brans and R.~H.~Dicke,
        Phys. Rev.
	{\bf 124}, 925-35 (1961).
\bibitem{weinberg} S.~Weinberg, {\em Gravitation and 
	Cosmology}, (Wiley, New York, 1972).
\bibitem{steve} S.~L.~Liebling, ``Massless Scalar Field Collapse In Brans-Dicke Theory'',
        M.~A.~Thesis, The University of Texas at 
        Austin (unpublished) (1995).
\end{references}
\end{document}